\documentstyle[12pt,aps]{revtex}
\begin{document}

%*************************************************
%************** D E F I N I T I O N S ************
%**************                       ************
\def\lsco{La$_{2-x}$Sr$_{x}$CuO$_{4}$\ }
\def\ybco{YBa$_{2}$Cu$_{3}$O$_{7-y}$\ }
\newcommand{\lapprox}{\stackrel{<}{\scriptstyle \sim}}
\newcommand{\gapprox}{\stackrel{>}{\scriptstyle \sim}}

\title{
\vspace{2cm}
\large
\bf 
Spin gap and magnetic coherence in a clean high-\mbox{\boldmath $T_{c}$}
superconductor \\
\vspace{0.8cm}  }

\author{B.\ Lake$^{1,2,3}$, G.\ Aeppli$^{4,3}$, T.\ E.\ Mason$^{1,2}$,
A.\ Schr\"{o}der$^{5}$, D.\ F.\ McMorrow$^{3}$, \\ 
K.\ Lefmann$^{3}$, M.\ Isshiki$^{6}$, M.\ Nohara$^{6}$, H.\
Takagi$^{6}$, S.\ M.\ Hayden$^{7}$.}

\address{
$^{1}$Department of Physics, University of Toronto, Toronto, ON M5S 1A7,
Canada \\
$^{2}$Oak Ridge National Laboratory,
Oak Ridge, Tennessee 37831, U.S.A.  \\
$^{3}$Department of Condensed Matter Physics and Chemistry, Ris\o\
National Laboratory, 
4000 Roskilde, Denmark \\
$^{4}$N.E.C Research, 4 Independence Way, Princeton, New Jersey 08540,
U.S.A. \\
$^{5}$Department of Physics, University of Karlsruhe, D-76128 Karlsruhe,
Germany \\
$^{6}$Institute for Solid State Physics, University of Tokyo, Roppongi
7-22-1, Minato-ku, Tokyo 106-8666, Japan.  \\
$^{7}$H.\ H.\ Wills Physics Laboratory, University of Bristol, 
Bristol BS8 1TL, U.K.}

\date{Nature, July 1, 1999}

\maketitle

\newpage

{\bf 
A notable aspect of high-temperature superconductivity in the copper oxides
is the unconventional nature of the underlying paired-electron state. 
A direct manifestation of the unconventional state is a pairing energy 
- that is, the energy required to remove one electron from the superconductor
- that varies (between zero and a maximum value) as a function of momentum 
or wavevector \cite{photo,ding}: the pairing energy for conventional 
superconductors is wavevector-independent \cite{miyake,scala}. The 
wavefunction describing the superconducting state will include not only the 
pairing of charges, but also of the spins of the paired charges. Each pair
is usually in the form of a spin singlet \cite{bardeen}, so there will also 
be a pairing energy associated with transforming the spin singlet into the 
higher energy spin triplet form without necessarily unbinding the charges. 
Here we use inelastic neutron scattering to determine the 
wavevector-dependence of spin pairing in \lsco, the simplest high-temperature
superconductor. We find that the spin pairing energy (or 'spin gap') is 
wavevector independent, even though superconductivity significantly alters 
the wavevector dependence of the spin fluctuations at higher energies.}

The experimental technique that we use is inelastic neutron scattering, 
for which the cross-section is directly proportional to the magnetic 
excitation spectrum and can be used to probe it as a function of 
wavevector and energy transfer. In addition, we have selected \lsco, 
the simplest of the high-temperature (high-$T_{c}$) superconductors. 
The material consists of nearly square CuO$_{2}$ lattices with Cu atoms 
at the vertices and O atoms on the edges alternating with LaSrO charge 
reservoir layers. In the absence of  Sr doping, the compound is an
antiferromagnetic insulator, where the spin on each Cu$^{2+}$ ion is
antiparallel to those on its four nearest neighbours. Because of the 
unit cell doubling, magnetic Bragg refections appear at wavevectors such 
as ($\frac{1}{2}$,$\frac{1}{2}$) (sometimes called ($\pi$,$\pi$) in the 
two-dimensional reciprocal space of the CuO$_{2}$ planes \cite{vaknin}). 
Doping yields a superconductor without long-range magnetic order but 
which has low-energy magnetic excitations peaked at the quartet of wavevectors
$\mbox{\boldmath $Q$}_{\delta}=(\frac{1}{2} ( 1 \pm \delta ), \frac{1}{2})$ 
and $( \frac{1}{2}, \frac{1}{2} ( 1 \pm \delta ) )$, shown in Figure 1a. 
The recent discovery of nearly identical fluctuations in the high-$T_{c}$ 
\ybco bilayer materials \cite{ybco} clearly indicates their relevance to 
the larger issue of high-$T_{c}$ superconductivity and validates the 
continued study of \lsco as the cuprate with the least structural and 
electronic complexity. 

The samples are single crystal rods grown in an optical image furnace. 
The most reliable measure of the quality of  bulk superconductors 
is the specific heat $C$. For our samples, there is a jump of 
$\Delta C / k_{B} T_{c}$ = 7 mJ/moleK$^{2}$ at $T_{c}$ = 38.5 K. As 
$T \rightarrow 0$, $C$ = $\gamma_{S} T$ where $\gamma_{S}$ is proportional 
to the electronic density of states at the Fermi level and has the value
$\gamma_{S} <$ 0.8 mJ/moleK$^{2}$. This together with an estimate of 
10 mJ/mole$K^{2}$ for the corresponding normal state $\gamma_{N}$ indicates 
that the bulk superconducting volume-fraction 1-$\gamma_{S}$/$\gamma_{N}$ 
of our samples is greater than 0.9. This, as well as the high value of 
$T_{c}$ and the narrowness of the transition, is evidence for the very 
high quality of our large, single crystals. The basic experimental 
configurations are similar to those employed previously 
\cite{previous1,previous2}. Figure 1a shows the reciprocal space regions 
probed. A series of scans like those indicated in the figure, performed for a 
range of energy transfers were used to build up the $\mbox{\boldmath $Q$}$-$E$ 
maps in Figure 1b and c, which show the scattering around the incommensurate 
peaks in the normal and superconduting states (here $E$ is the energy 
transfer).

Figure 1b shows that the normal state excitations at 38.5 K
are localized near $\mbox{\boldmath $Q$}_{\delta}$ but are entirely
delocalized in $E$. In other words the magnetic fluctuations which are 
favoured are those with a particular spatial period $1/\delta$ 
corresponding to $\mbox{\boldmath $Q$}_{\delta}$, but no particular 
temporal period. Cooling below $T_{c}$ produces a very different image 
in $\mbox{\boldmath $Q$}$-$E$ space. In Figure 1c all low frequency 
excitations ($E \leq 5$ meV) seem to be eliminated and there is an 
enhancement of the signal above 8 meV at the incommensurate wavevectors. 
The signal now has obvious peaks at around $E$=11 meV and 
$\delta=0.29 \pm 0.03$ reciprocal lattice units (r.l.u.). We can thus 
visualize the zero point fluctuations in the superconductor as magnetic 
density waves undergoing (damped) oscillation with a frequency of 2.75 THz. 
In the normal paramagnetic state, the motion of the density waves becomes 
entirely incoherent.  

Figure 2a-2c shows a series of constant-$E$ cuts through the data in 
Figure 1. These graphs demonstrate that superconductivity induces a complete 
loss of signal for $E$=2 meV (2a), a significant intensity-preserving 
sharpening of the incommensurate peaks for $E$=8 meV (2b), and a large 
enhancement of the peaks for $E$=11 meV (2c). The peak narrowing in (2b) 
and (2c) corresponds to a spectacular superconductivity-induced rise in the 
magnetic coherence lengths (defined as the resolution-corrected inverse
half-widths at half-maxima obtained as in \cite{thom1}) from 
$20.1 \pm 0.9$ \mbox{ \AA} to $33.5 \pm 2.0$ \mbox{ \AA} and $ 25.5 
\pm 0.1 $ \mbox{ \AA} to $ 34.3 \pm 0.8 $ \mbox{ \AA}, respectively. 

Figure 3a-c displays constant-$\mbox{\boldmath $Q$}$ spectra both away
from $\mbox{\boldmath $Q$}_{\delta}$ (Figure 3a and b) and at 
$\mbox{\boldmath $Q$}_{\delta}$ (Figure 3c). Superconductivity removes the 
low-$E$ signal below a threshold energy, while it enhances the higher-$E$ 
signal close to $\mbox{\boldmath $Q$}_{\delta}$. The threshold for 
$T < T_{c}$ appears the same for the three wavevectors shown in Figures 3a-c, 
with the increase in intensity first visible in all cases at 6 meV. To 
quantify how superconductivity changes the spectra, we fit the data with the 
convolution of the instrumental resolution (full-width-half-maximum = 2 meV) 
and
\begin{equation}
S(\mbox{\boldmath
$Q$},E)=\frac{1}{1-exp(E/k_{B}T)}\frac{AE'\Gamma}{\Gamma^{2}+E^{2}}
\end{equation}
where
\begin{equation}
E' = Re \left\{ [ ( E-\Delta+i\Gamma_{s} ) ( E+\Delta+i\Gamma_{s} ) ]^{1/2} \right\}
\end{equation}
and $A$ is the amplitude, $\Delta$ is the spin gap, $\Gamma$ is the inverse 
lifetime of spin fluctuations with $E\gg\Delta$ (if $\Delta\ll\Gamma$),
$E'$ is an odd function of $E$ which defines the degree to which the
spectrum has a gap and $\Gamma_{s}$ is the inverse lifetime of the 
fluctuations at the gap edge. 

In the normal state, the best fits are obtained for $\Delta=0$ meV, and the 
fitted value of $\Gamma$ is essentially $\mbox{\boldmath $Q$}$-independent, 
(Figure 3d). Thus, the lower-amplitude fluctuations with wavevectors different
from $\mbox{\boldmath $Q$}_{\delta}$ have lifetimes similar to those at the 
incommensurate peak positions. The $\mbox{\boldmath $Q$}$-dependence of the 
signal is entirely accounted for by the $\mbox{\boldmath $Q$}$-dependence of 
the real part $\chi'(\mbox{\boldmath $Q$})$ of the  magnetic susceptibility 
(Figure 3e) which, when $\Delta=0$, is simply the amplitude $A$. In the 
superconducting state, $\Gamma$ (Figure 3d), which characterizes the shape 
of the spectrum well above the spin gap, becomes strongly 
$\mbox{\boldmath $Q$}$-dependent. At the same time, 
$\chi'(\mbox{\boldmath $Q$})$ (Figure 3e), related via a 
Kramers-Kronig relation to the parameters in Equation (1), is suppressed. 
This explicitly demonstrates that superconductivity reduces the tendency 
towards static incommensurate magnetic order in \lsco.

Figure 4 shows the $\mbox{\boldmath $Q$}$ dependence of the spin gap $\Delta$. 
As anticipated from inspection of the data in Figure 3, $\Delta$ is 
$\mbox{\boldmath $Q$}$-independent and has the value 6.7 meV. The gap is quite
sharp for our sample, with $\Gamma_{s}\leq0.2$ meV for all 
$\mbox{\boldmath $Q$}$. Also shown in Figure 4 are the results for $x$=0.15 
\cite{yamada,petit} and 0.14 \cite{thom1}. 
$\Delta(\mbox{\boldmath $Q$}_{\delta})$ is indistinguishable for the present 
$x$=0.163 and the older $x$=0.14 samples; the difference in the low-$E$ 
behavior is primarily due to the much larger damping 
($\Gamma_{s}=1.2$ meV for $x$=0.14 \cite{thom2}). In addition, the 
$\mbox{\boldmath $Q$}$-independence of $\Delta(\mbox{\boldmath $Q$})$ 
is consistent with the $\mbox{\boldmath $Q$}$-independent but
incomplete suppression of the magnetic fluctuations in the $x$=0.14 sample 
\cite{thom93}. In contrast, the results of \cite{yamada} and \cite{petit} 
show a large discrepancy with $x$=0.163, where the spin gap quoted in these 
papers is defined as the threshold for visible scattering. Nevertheless the 
results of \cite{yamada} are consistent with our work if we use the 
definition - advocated here and in \cite{thom2} - of $\Delta$ given by 
Equation (2). Fitting the data of \cite{yamada} to Equation (1) with 
$\Delta$= 6.7 meV yields $\Gamma_{s}$ = 0.5 meV, a value intermediate 
between our findings of 1.2 and 0.1 meV for $x$=0.14 and 0.163.

Our experiments show that superconductivity produces strongly
momentum-dependent changes in the magnetic excitations with energies
above a momentum-independent spin gap. The data in their entirety do not 
resemble the predictions 
\cite{theory1,theory2,theory3,theory4,theory5,theory6} for
any superconductors, be they $s$-wave or $d$-wave.  Most notably, all 
$d$-wave theories anticipate dispersion in the spin gap which would 
have been observed over the wavevector range and for the energy 
resolution of the present experiment.  At the same time, $s$-wave 
theory cannot account for the value of the spin gap. We are 
unaware of calculations which yield the dramatic incommensurate peak
sharpening and enhancements seen above the spin gap, while at the same 
time showing a large reduction in the real part of the magnetic 
susceptibility.

There are other difficulties with the conventional weak-coupling 
$d$-wave approach which posits nodes and therefore a smaller relative 
superconductivity-induced reduction in scattering between rather than 
at the incommensurate peaks. Figure 2b shows the opposite - just above 
the gap energy, the incommensurate peak intensities are preserved while 
the scattering between the peaks is suppressed. Furthermore, the peak 
sharpening in momentum space for $\hbar\omega > \Delta$ finds a precedence 
only in quantum systems, such as $S$=1 antiferromagnetic (Haldane) spin 
chains and rotons in superfluid helium, which have well-defined gaps with 
non-zero minima. Thus, while our statistics and resolution cannot exclude 
a small population of spin-carrying subgap quasiparticles, the systematics 
of the signal found near the gap energy make such quasiparticles improbable.
As for any other spectroscopic experiment, we can only place an upper 
bound on the signal below the dispersionless gap. Inspection of Figure 3 
shows that in between the incommensurate peaks at $\mbox{\boldmath $Q$}
=(\frac{1}{2}(1+\frac{\delta}{2}),\frac{1}{2}(1-\frac{\delta}{2}))$,  
where ordinary weak-coupling $d$-wave theories generally anticipate nodes 
in the spin gap, the intensity for 2 meV at 5 K is less than 14 \% of 
what was seen at $T_{c}$ and below 5 \% of that observed for the 
incommensurate peaks at 5 K.

Given the overwhelming evidence for $d$-wave superconductivity in the 
hole-doped high-$T_{c}$ superconductors \cite{photo,ding,tsuei,chen}, we 
see our data not as evidence against $d$-wave superconductivity but as proof 
that the spin excitations in the superconducting state do not parallel the 
charge excitations in the fashion assumed for ordinary $d$- and $s$-wave 
superconductors. Our measurements, which are sensitive exclusively to the 
spin sector, taken together with the evidence for $d$-wave superconductivity 
in the charge sector suggest that the high-$T_{c}$ superconductors are 
actually Luther-Emery liquids, namely materials with gapped (triplet) 
spin excitations and gapless spin zero charge excitations \cite{emery1,Rokh}. 
Luther-Emery liquids arise in one-dimensional interacting Fermi systems, 
which formally resemble two-dimensional $d$-wave superconductors - the 
dimensionality (zero) of the nodal points where the gap vanishes in the 
two-dimensional copper oxide is the same as that of the Fermi surface of a 
one-dimensional metal. There are other arguments for the applicability of the 
concept of Luther-Emery liquids. The first is that theory indicates 
that such liquids are the ground states of ladder compounds, one-dimensional 
strips of finite width cut from CuO$_{2}$ planes 
\cite{emery3,dagotto,rice,emery5}. The second involves the break-down of  
spin-charge separation when the spin gap collapses to zero, which can be 
brought about by a magnetic field whose Zeeman energy matches the spin 
gap energy. The 6.7 meV spin gap which we measure is much closer to the 
Zeeman energy of the upper critical field measured \cite{andoh} for samples 
similar to ours than to an ordinary Bardeen-Cooper-Schrieffer pairing 
energy $\geq 3.5 k_{B}T_{c} = 11.6$ meV.

We thank K.\ N.\ Clausen for his help and support during the experiments, 
and B.\ Batlogg, G.\ Boebinger, V.\ Emery, K.\ Kivelson, H.\ Mook, D.\ Morr, 
D.\ Pines, Z-X.\ Shen, C-C.\ Tsuei and J.\ Zaanen for helpful disussions.  
Work done at the University of Toronto was sponsored by the Natural Sciences 
and Engineering Research Council and the Canadian Institute for Advanced
Research, while work done at Oak Ridge was supported by the US DOE. 
TEM acknowledges the financial support of the Alfred P.\ Sloan 
Foundation, and AS acknowledges the assistance of the TMR program.

Correspondence and requests for materials should be addressed to BL 
(e-mail: bella@phonon.ssd.ornl.gov).

\newpage

\begin{figure}
\caption{
The reciprocal space regions over which measurements were made and the 
resulting data as a function of wavevector and energy transfer. {\bf a} 
is a reciprocal space diagram of the CuO$_{2}$ planes in \lsco where 
the black circles represent the incommensurate peaks surrounding 
($\pi$,$\pi$). The coloured strips give the data collected at an energy 
transfer of 9 meV and temperature of 5 K and show the areas probed in 
this experiment. The lower strip passes through two of the peaks and 
provides the signal, whereas the upper strip lies far from the peaks 
and is used as the background. {\bf b}, is a $\mbox{\boldmath $Q$}$-$E$ 
map measured at the transition temperature of 38.5 K, for the wavevectors 
shown in {\bf a} and energy transfers from 2 to 16 meV, the parameter $h$ 
defines the two-dimensional wavevector 
$\mbox{\boldmath $Q$}$=[0.56$h$,0.44$h$]. The data displayed are background 
subtracted and the thermal population factor $(exp(-E/k_{B}T)-1)^{-1}$ has 
been divided out to give the quantity $S$($\mbox{\boldmath $Q$}$,$\omega$). 
The colouring of the squares indicates the intensity observed in units of 
counts per 15 minutes. {\bf c}, A similar map to {\bf b} but for the 
superconducting phase at 5 K. The sample used for these measurements 
consisted of five crystals grown by the travelling solvent floating zone 
method; each was approximately 4 mm in diameter and 20 mm long. In order 
to maximise signal the crystals were mounted on a single holder so that 
their axes were parallel to within 0.8 degrees. The measurements were 
performed on the new RITA spectrometer at Ris\o\ National Laboratory 
\protect\cite{rita}. RITA differs from its predecessor, TAS6, by use of a velocity 
selector for better filtering of the incident beam, superior beam optics 
between the reactor and the sample and a large position sensitive detector. 
In this experiment pixels at different vertical heights on the detector 
were binned separately so that a single instrumental scan produced a 
two-dimensional plot in reciprocal space as seen in {\bf a}.
}
\end{figure}

\begin{figure}
\caption{
Constant-$E$ scans at various energies through the incommensurate peaks 
at $T_{c}$ and in the superconducting phase at 5 K. {\bf a}, The data for 
2 meV well below the spin gap energy of 6.7 meV. The incommensurate peaks 
are clearly visible at $T_{c}$ but are completely suppressed in the 
superconducting phase. {\bf b}, scattering at 8 meV, somewhat above the 
gap energy. The peak amplitudes are approximately equal at the two 
temperatures while their lineshapes are different with scattering at 5 K 
narrower than at 38.5 K. {\bf c} The scattering at 11 meV well above the 
spin gap. The scattering amplitude at 5 K is significantly greater than at 
38.5 K. The solid lines through the data were obtained by following the 
analysis procedures of \protect\cite{thom1}.
}
\end{figure}

\begin{figure}
\caption{
Spectra at various wavevectors, and the $\mbox{\boldmath $Q$}$-dependence of 
the inverse lifetime and susceptibility extracted by fitting such profiles. 
(The wavevectors range from between the incommensurate peaks up to the peak 
maxima). Panel {\bf  a} shows the constant-$\mbox{\boldmath $Q$}$ spectrum 
at h=1.000 (a wavevector exactly in between the two peaks), {\bf  b} shows the 
spectrum at h=1.095 (a position on the inner slope of the peak) and 
{\bf  c} gives the spectrum at h=1.135 (the peak maximum). The spin gap 
is present at all three wavevectors in the superconducting phase with the 
scattering eliminated below 6 meV. As h is changed from 1.000 to 1.135 
there is a clear maximum in the response at energies around 11 meV in the 
superconducting phase. {\bf  d} and {\bf e} show the values of the energy 
scale $\Gamma$ and the real part of the magnetic susceptibility $\chi '$ 
(in units of counts per 15 minutes) as functions of wavevector for both 
38.5 K and 5 K. These quantities were extracted by fitting to the data the 
lineshape given in Equation (1) convolved with the instrumental resolution. 
$\Gamma$ is approximately constant at $T_{c}$ but strongly 
$\mbox{\boldmath $Q$}$-dependent at 5 K and $\chi '$ is smaller at 5 K than 
at 38.5 K although it has a similar wavevector-dependent profile at the two 
temperatures.
}
\end{figure}

\begin{figure}
\caption{ 
The wavevector dependence of the spin gap in the superconducting state 
at 5 K. The gap was extracted by fitting Equation (1), 
convolved with the instrumental resolution, to
constant-$\mbox{\boldmath $Q$}$ spectra such as those shown in Figure
3a-c. The gap values are given by the open circles and are independent 
of wavevector. The solid line gives the fitted wavevector-independent 
value of the gap which is $\Delta$=6.7 meV. The solid symbols show the 
spin gaps determined previously for \lsco. The filled square gives the 
gap for an $x$=0.14 sample, the gap was extracted by the same method as 
used in this paper and found to be of a similar size. The filled triangle 
and filled diamond give the gap for two different $x$=0.15 samples; in 
both cases the gap was defined as the threshold for visible scattering 
without considering resolution broadening or damping at the gap edge. 
Using this definition the gap was found to have a much smaller value.  
In all except the present work, the spin gap was established at only a 
single wavevector.
}
\end{figure}

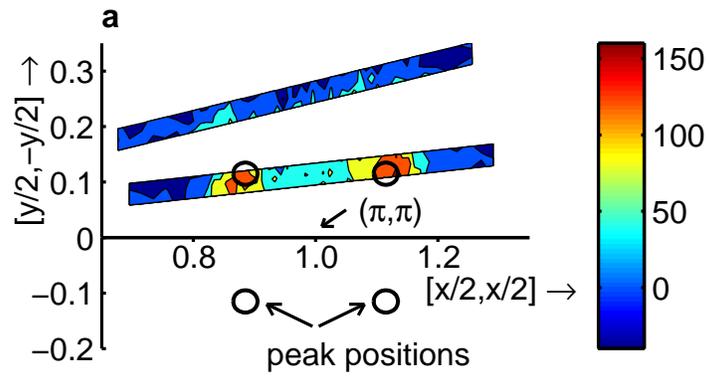
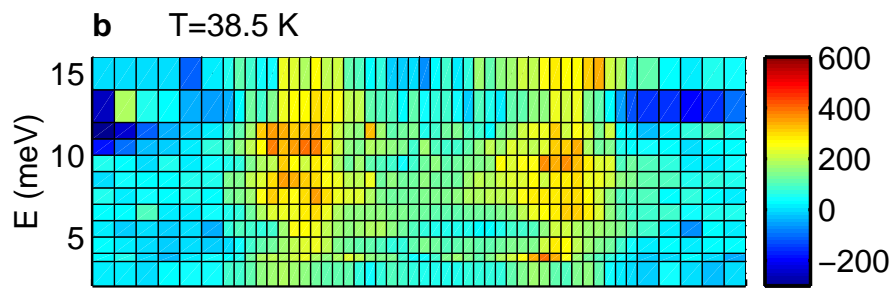
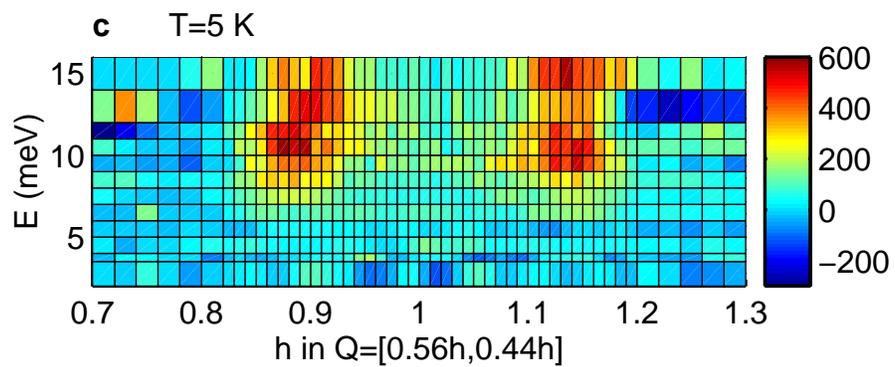
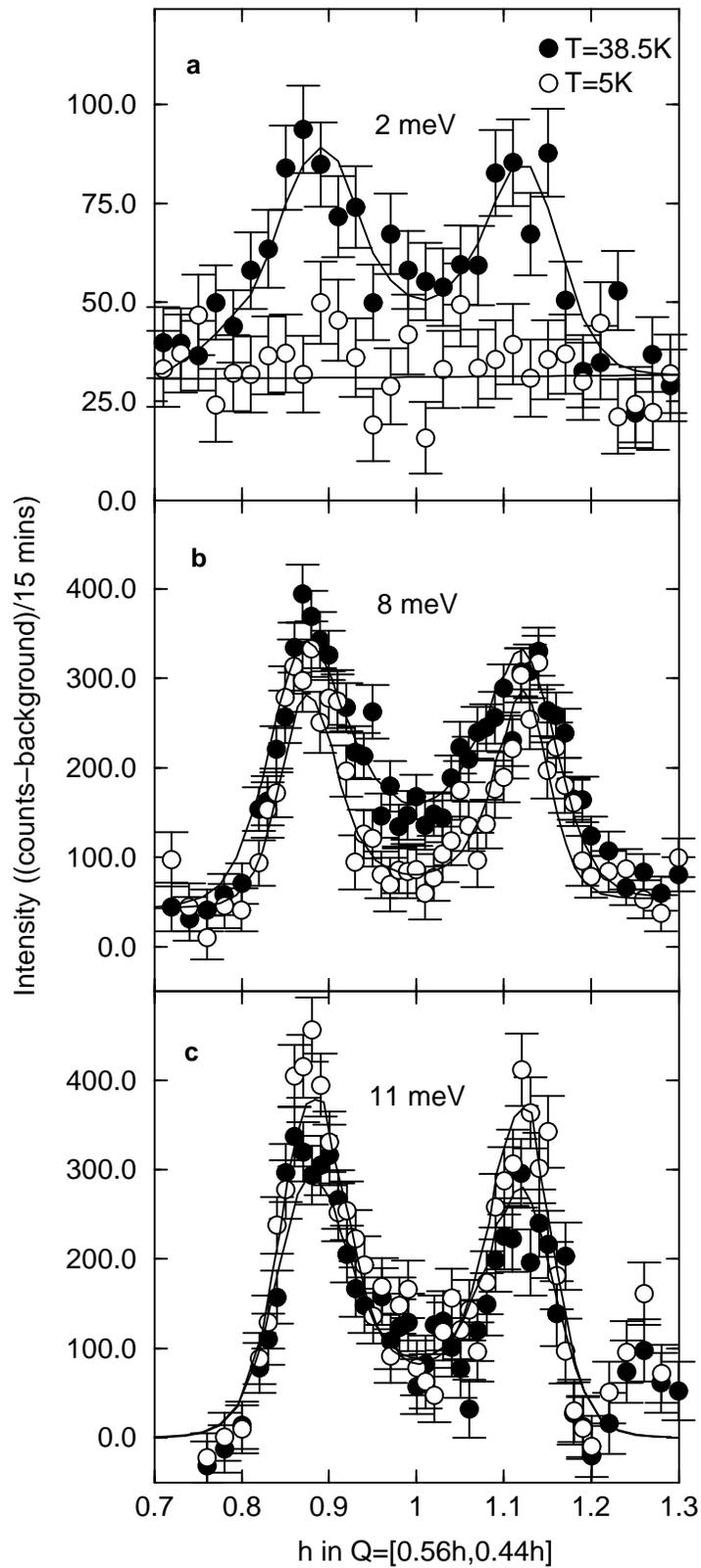
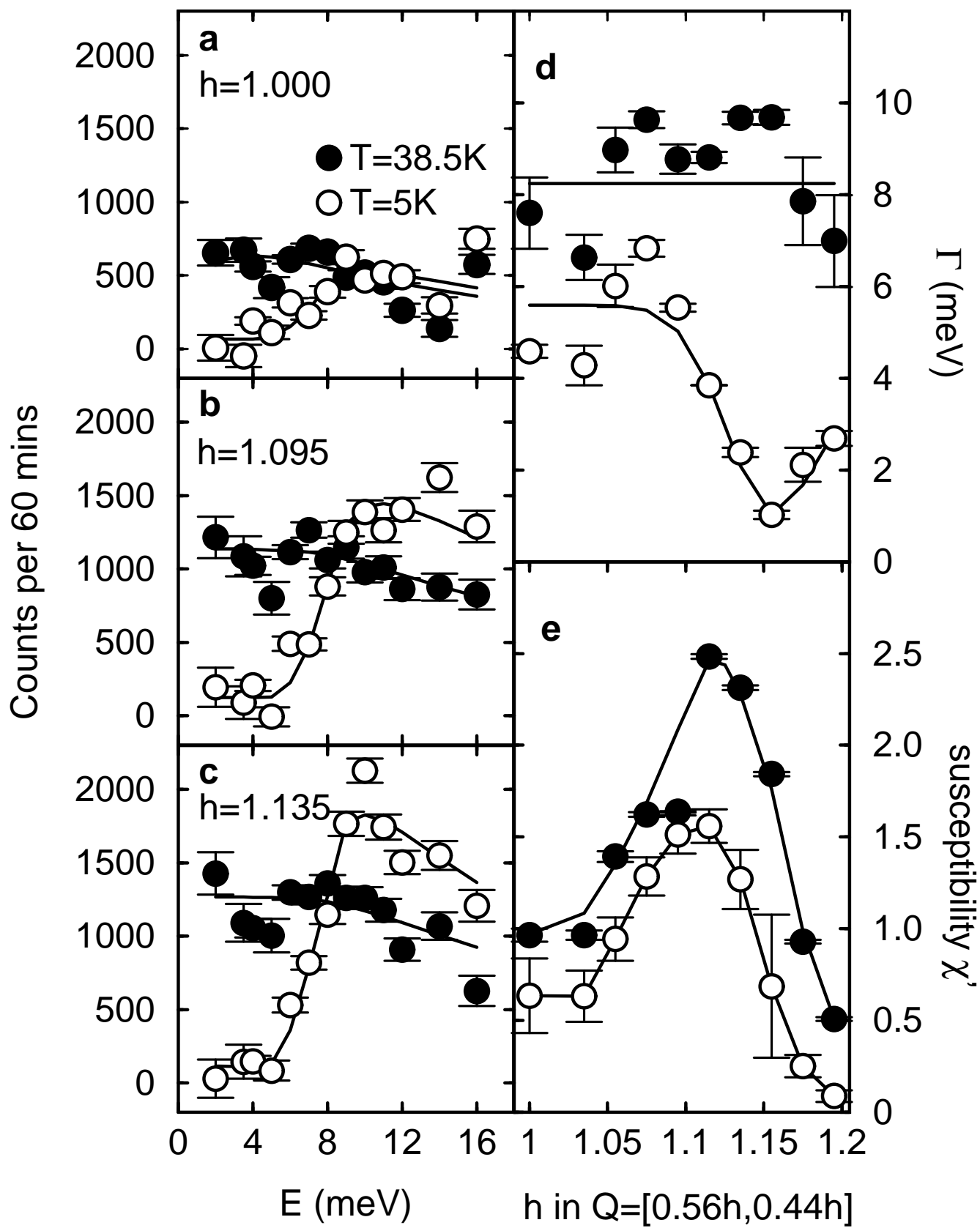
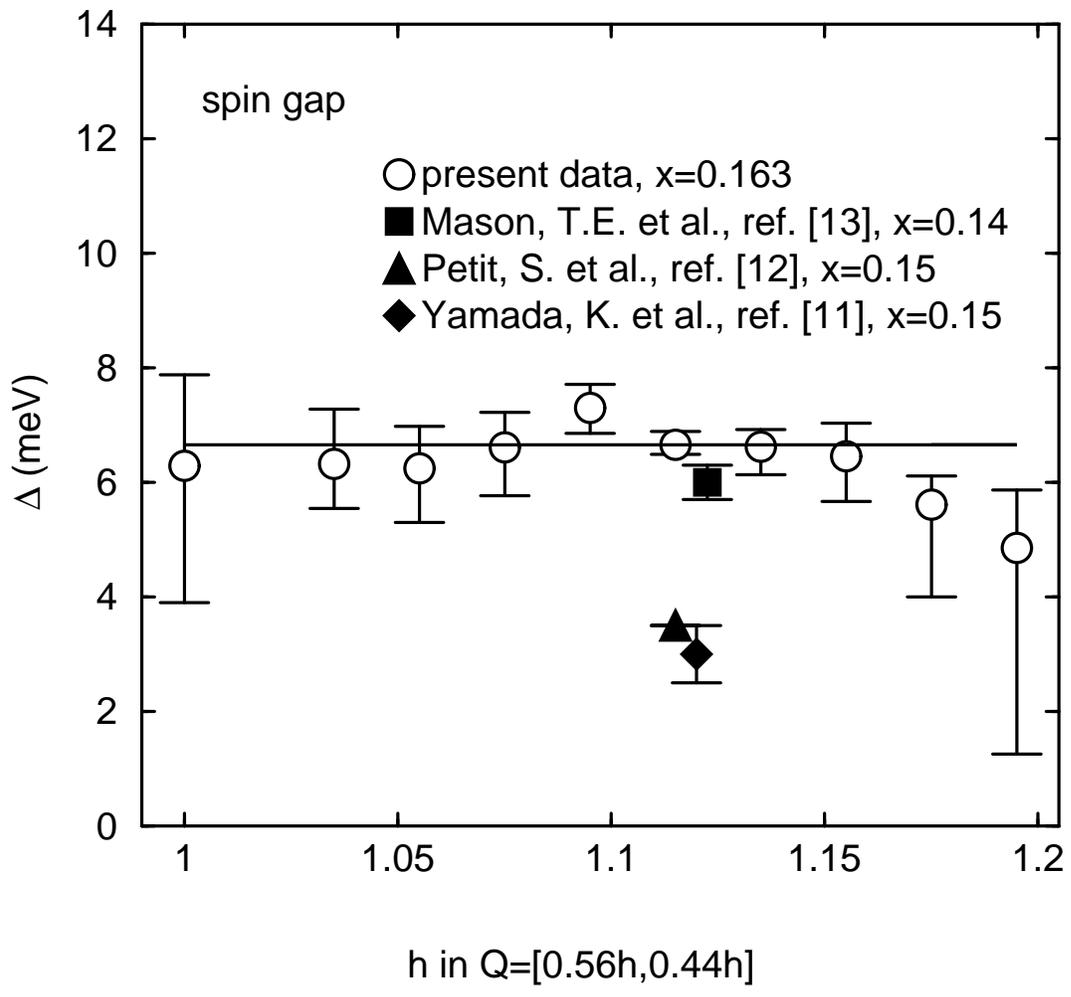
\end{document}